\newcommand{\beq}{\begin{equation}}
\newcommand{\enq}{\end{equation}}
\newcommand{\bee}{\begin{eqnarray}}
\newcommand{\ene}{\end{eqnarray}}
\newcommand{\bem}{\begin{mathletters}}
\newcommand{\enm}{\end{mathletters}}

\documentstyle[eqsecnum,aps]{revtex}

\begin{document}
\draft
\title{Solutions of Gross--Pitaevskii equations beyond the
hydrodynamic approximation: Application to the vortex problem.}
\author{Vladimir V. Konotop\footnote{Electronic address:
konotop@alf1.cii.fc.ul.pt}}

\address{Departamento de F\'{\i}sica e
Centro de F\'{\i}sica da Mat\'eria Condensada,  Universidade de Lisboa,
Complexo Interdisciplinar, \\ Av.
Prof. Gama Pinto, 2, Lisbon, P-1649-003 Portugal}

\author{V\'{\i}ctor M. P\'erez-Garc\'{\i}a\footnote{Electronic address:
vperez@ind-cr.uclm.es}}

\address{Departamento de Matem\'aticas, E.T.S.I. Industriales, Universidad de Castilla-La Mancha,
\\ Avda. Camilo Jos\'e Cela, 3, Ciudad Real, 13071 Spain.}

\date{\today}
\maketitle

\begin{abstract}
We develop the multiscale technique to describe excitations
of a Bose-Einstein condensate (BEC) whose characteristic scales are
comparable with the healing length,  thus going beyond the
conventional hydrodynamical approximation. As an application of the
theory we derive approximate explicit
 vortex and other solutions. The {\em dynamical stability}
 of the vortex is discussed on the basis of
the mathematical framework developed here, the result being that its stability is
granted at least up to times of the order of seconds, which is the condensate lifetime.
Our analytical results are confirmed by the numerical simulations.

\end{abstract}


%\narrowtext

\section{Introduction}

 Since the remarkable experimental realization \cite{Science} of the
Bose-Einstein condensation (BEC) there has been an explosion of
the experimental
 and theoretical activity devoted to the physics of dilute ultracold bosonic
gases \cite{web}.

 From the theoretical point of view, many of current condensate systems
are cold enough that
the mean field theories can be applied to describe the ground state of
the system  \cite{Griffin}. The most classical model arising in this
 context for the order parameter of the system is the
the Gross-Pitaevskii (GP) equation.  In fact, many relevant
features of the condensate such as free expansion \cite{free}, low
energy excitations \cite{lowenergy}, collapse
\cite{collapse,Humberto}, dynamics of multicomponent systems
\cite{multi}, vortex generation and dynamics
\cite{Nature,vortices}, and many others, can be understood within
the framework of this ``simplest" mathematical model.

Mathematically, the GP equation, to be written
explicitly below, is an equation of Nonlinear Schr\"odinger type.
These equations have been studied extensively both in the physical and
 mathematical literatures \cite{Vazquez,Sulem} since they provide an
universal model for the study of the dynamics of envelope waves.
Some fields where these equations arise are plasma physics
\cite{plasma}, fiber and integrated optics \cite{Hasegawa}, wave
propagation in Kerr media \cite{Newell}, water waves \cite{Dodd},
biomolecule dynamics \cite{Davidov} and fundamentals of quantum
mechanics \cite{Rosales94}. One of the distinctive features of the
equation as it appears in BEC problems is the presence of an
external potential -the trap-, which affects essentialy the
spectrum of the ``linear part", i.e. of the underlying
Schr\"{o}dinger equation.

The mathematical analysis of the GP equation in the situations
common in BEC is very complicated and this is why most theoretical
studies deal with either numerical simulations or some kind of
approximations such as the Thomas--Fermi, the hydrodynamic
approach or simply reduce   the number of degrees of freedom by
using the variational approach  based on trial functions. To be
precise let us write explicitly  the equation for the order
parameter, whose form is
\begin{equation}
\label{GP}
i\hbar \frac{\partial \Psi}{\partial t}=-\frac{\hbar^2}{2m}\nabla^2\Psi
+V_{ext}({\bf r})\Psi +g|\Psi|^2\Psi,
\end{equation}
where the external potential can be well approximated by
\begin{equation}
\label{poten}
V_{ext}=\frac{1}{2} m \left(\omega_x^2x^2+\omega_y^2y^2+\omega_z^2z^2\right),
\end{equation}
and $g=4\pi\hbar^2a_s/m$.

The simplest approximation that can be done to obtain information on the
{\em ground state}
is the so called Thomas--Fermi approximation,  which
corresponds to  neglecting the kinetic energy term under the
assumption that it
 is much less than the self-interaction energy. In the cases when the
trap is smooth enough this approximation is not bad although it is not consistent, since the assumption of zero kinetic energy provides a
profile that has an infinite value of that quantity.

The following level of approximation, the hydrodynamic approach, can be explained by
introducing new variables $n({\bf r}, t)$ and ${\bf v}({\bf r},t)$
through the relations
\begin{mathletters}
\label{new}
\begin{eqnarray}
\Psi({\bf r},t) & = & \sqrt{n({\bf r},t)}e^{iS({\bf r},t)}, \\
{\bf v}({\bf r},t) & = & \frac{\hbar}{m}\nabla S({\bf r}, t).
\end{eqnarray}
\end{mathletters}
Then, one arrives to the system of equations
\beq
\label{hyd2}
\partial_t n+\nabla\cdot ({\bf v} n)=0,
\enq
\beq
\label{hyd3}
m\partial_t{\bf v}+\nabla\left(V_{ext}+g
n-\frac{{\hbar}^2}{2m\sqrt{n}}\nabla^2\sqrt{n}+\frac{mv^2}{2}
\right)=0.
\enq
The conventional hydrodynamic approximation corresponds to neglecting the
term proportional to $\nabla^2\sqrt{n}$ under the assumption that the
density is a slowly varying function \cite{Dalfovo}. For a spherically symmetric potential
 ($\omega_x=\omega_y=\omega_z=\omega$) this can be done in the limit
$Na_s/a_0\gg 1$, where
$a_s>0$ is the scattering length and $a_0=\sqrt{\hbar/m\omega}$
is the harmonic oscillator length, i.e. the trap size.

To the best of authors knowledge the {\em mathematical} limits of validity
of the above mentioned approximations have not been established
(which would imply study of the next orders of the expansion or making
appropriate rigorous estimates), so far.
 In particular, using the estimate $n\sim 10^{12}$ cm$^{-3}$ one finds
that for $a_s\sim 10$ nm the term with $\nabla^2\sqrt{n}/\sqrt{n}$
is much less than $gn$ for excitations whose characteristic length
$\Lambda$ is of order of $10^{-3}$ cm. Such a $\Lambda$ is
comparable with the characteristic size $a_0$ of the trapping
potential.  Thus the question about analytical description of the
excitations which vary on scales much less than $a_0$ appears. It
is important to notice that in those cases the terms
$\frac{\hbar^2}{m}\frac{\nabla^2\sqrt{n}}{\sqrt{n}}$ and $gn$
become comparable with each other and have to be treated in
equivalent manner. Using the wave terminology this situation can
be characterized as balance of dispersive and nonlinear effects,
which should lead to stable spatially localized excitations.

 It is our intention in this paper to describe an analytical tool,
{\em the multiscale approximation}, which is able to provide
approximate solutions of the problems under consideration in explicit
analytical form. In less rigorous
 terms this technique has been applied in the context of BEC in
 Refs.~\cite{Humberto,Henar} to problems with negative scattering length.
 The application to positive scattering length presents some new
characteristics as we will see later (in particular it implies
different scaling order). Specifically, we will consider a BEC in
a parabolic cilindrically or spherically symmetric confining
potential and obtain a set of background solutions in the form of
{\em black holes} or {\em vortices} whose characteristic spatial
scale is much larger than the range of inter-atomic interactions
but smaller than the trap size. Working in this way we will be
able not only to obtain an approximate analytical shape for the
vortex solutions but also to provide results on their dynamical
stability, a point which is still of great interest after the many
studies devoted to it \cite{Fetter,Juanjo}.

The outline of the paper is as follows. First, in Sec.
 \ref{statement} we pose the problem in precise terms.
  As a second step, we apply the technique of multiscale
  expansion to our particular problem in Sec. \ref{multiscale}.
   In Sec. \ref{stability} we
 obtain a particular set of solutions and
discuss their stability. Finally, in Sec. \ref{conclussions}, we summarize our conclusions.

\section{Statement of the problem}
\label{statement}

We start with the three-dimensional GP equation in the form
\begin{equation}
\label{GPnoloc}
i\hbar \frac{\partial \Psi}{\partial t}=-\frac{\hbar^2}{2m}\nabla^2\Psi
+V_{ext}({\bf r})\Psi +J_\lambda({\bf r})\Psi.
\end{equation}
We will preserve the integral form of the interaction term [c.f. Eq.
(\ref{GP})]
\begin{eqnarray}
\label{nonloc}
J_{\lambda} ({\bf r})=g \int
 V_\lambda ({\bf r}, {\bf r}')
 |\Psi({\bf r}',t)|^2\,d{\bf r}',
\end{eqnarray}
since there are situations where its role could be important
\cite{colapsono} and it does not complicate the formalism. Here
$V_{\lambda}({\bf r}, {\bf r}')$ describes two-body interactions
and $\lambda$ is the effective interaction length (mathematically,
the size of the kernel).

The wave function is normalized to the total number of particles $N$
\beq
\label{norm}
\int |\Psi({\bf r})|^2d{\bf r}=N.
\enq

It is reasonable to assume that the effective interaction length $\lambda$
parametrizes the kernel so that
\begin{equation}
\label{e1}
\lim_{\lambda\to 0}
 V_\lambda ({\bf r}, {\bf r}')
=\delta ({\bf r}-{\bf r}').
\end{equation}
The last limit corresponds to the usual case of local interactions and
gives rise to the $|\Psi|^2$ nonlinear term in the GP equation [i.e.
(\ref{GPnoloc}) transforms into (\ref{GP})].
Bearing in mind this last fact we can narrow the
class of the kernels under consideration by requiring
\begin{equation}
\label{}
 V_\lambda ({\bf r}, {\bf r}')\equiv
\frac{1}{\lambda^3} U \left(\frac{|{\bf r}-{\bf r}'|}{\lambda}\right),
\end{equation}
where
\begin{equation}
\int U(|{\tilde{\bf r}}-{\tilde{\bf r}}'|)\,d{\tilde{\bf r}}'=1.
\end{equation}
and ${\tilde{\bf r}}={\bf r}/\lambda$. As it is clear from the
definition of $\lambda$, it must be of order of the scattering length:
\beq
\label{lambda}
\frac{\lambda}{a_s}=O(1)
\enq
In what follows, however we keep $\lambda$ as a parameter.

We will concentrate on a BEC in a trap with  either cilindrical
(with $z$-axis being the symmetry axis) or spherical
symmetry, i. e. $\omega_\alpha=\omega$, where $\alpha=x,y$ in the
former case
and $\alpha=x,y,z$ in the last case. We define the characteristic
scales of the condensate in the trasversal, $a_0=\sqrt{\hbar/(m\omega)}$,
and longitudinal,
$a_{z}=\sqrt{\hbar/(m\omega_z)}$, directions.  In the case of spherical
symmetry these scales coincide, $a_z=a_0$.

In order to provide a formal description of the system it is
convenient to scale out the GP equation. To this end we introduce the
dimensionless time and wavefuntion through
\begin{mathletters}
\begin{eqnarray}
\tau & = & \frac{\hbar}{2m\lambda^2}t, \\
\psi({\tilde{\bf
r}}, \tau) & = & 2\lambda\sqrt{\frac{2\pi a_s}{N}}\Psi({\bf r}, t).
\end{eqnarray}
\end{mathletters}
Then the renormalized GP equation takes the form
\begin{equation}
\label{e2}
i \frac{\partial \psi}{\partial \tau}=-\Delta\psi
+\tilde{V}_{ext}(\tilde{{\bf r}})\psi +\tilde{J}(\tilde{{\bf r}})\psi
\end{equation}
where
\begin{eqnarray}
\tilde{J}(\tilde{{\bf r}})= \int U (\tilde{{\bf r}}-\tilde{{\bf r}}')
|\psi(\tilde{{\bf r}}',\tau)|^2\,d\tilde{{\bf r}}'\,,
\end{eqnarray}
\begin{eqnarray}
\label{ext}
\tilde{V}_{ext}(\tilde{{\bf r}})=\left\{\begin{array}{ll}
(\Omega^2\tilde{{\bf r}}_{\bot}^2+
\Omega_z^2\tilde{z}^2)  & \,\,\,\,\,\mbox{cilindrical symmetry} \\
\Omega^2\tilde{{\bf r}}^2
 & \,\,\,\,\,\mbox{spherical symmetry}
\end{array} \right.
\end{eqnarray}
$\tilde{{\bf r}}_\perp=(\tilde{x},\tilde{y})$, $\Omega=\lambda^2/a_0^2$ and
$\Omega_z=\lambda^2/a_{z}^2$.

Having scaled out the kernel which describes the nonlocal interactions
we
have to impose the condition that the characteristic scale of the kernel
variation is of order one, i. e.
\[
\Bigg| \frac{1}{U(\tilde{r})}\frac{\partial U(\tilde{r})}{\partial
\tilde{r}}\Bigg|=O(1).
\]
where  $\tilde{r}=|\tilde{{\bf r}}|$ and $\tilde{{\bf r}}$ must be
substituted by $\tilde{{\bf r}}_{\bot}$ in the case of the cilindrical
symmetry.

\section{Multiscale analysis.}
\label{multiscale}

 The multiscale technique is a mathematical technique for the analysis of problems where
there are different spatial scales involved in the solution. The
problem at hand possesses three important scales: the scattering
length $a_s$, the trap size $a_0$ and the healing length $\xi=(8\pi
na_s)^{-1/2}$. We will
 concentrate on the analysis of excitations of the superfluid corresponding
 to the
scale of the healing length, which occur on top of a background with
spatial scale of the order of the trap size. In other words our theory
will be valid for cases when the above three parameters are
related as follows
\begin{equation}
\label{paramet}
a_s\ll\xi\ll a_0.
\end{equation}

For the first step we recall the properties of the linear spectral
problem
\begin{equation}
\label{lin1}
L_d|n,l,m\rangle={\cal E}_{n,l,m}|n,l,m\rangle,
\end{equation}
where $L_d={\cal L}_d+{\cal M}_d$.  The operator ${\cal L}_d$ describes
the radial dependence
\begin{equation}
\label{lin2}
{\cal L}_d=-\frac{1}{\rho_0^{d-1}}\frac{\partial}{\partial
\rho}\rho_0^{d-1}
\frac{\partial}{\partial \rho_0}+\Omega^2\rho_0^2,
\end{equation}
and
\begin{mathletters}
\label{M}
\begin{eqnarray}
{\cal M}_2 & =& \frac{\partial^2}{\partial z^2}+
\frac{1}{\rho_0^2}\frac{\partial^2}{\partial \phi^2}, \\
{\cal M}_3 & = & \frac{1}{\rho_0^2\sin\theta}\frac{\partial}{\partial
\theta}\sin\theta \frac{\partial}{\partial \theta}+
\frac{1}{\rho_0^2}\frac{\partial^2}{\partial \phi^2}.
\end{eqnarray}
\end{mathletters}
The indexes $n$ and $m$ refer to radial and magnetic quantum numbers
while $l$ is the azimutal (in the spherical case) or longitudinal (in
the cilindrical case) quantum number (see Appendix). Hereafter $d=2$
and $d=3$ are associated with cilindrical and spherical symmetries.

For the next consideration it is convenient to single out the radial
part of the eigenfunctions and represent
\bee
\label{separ}
|n,l,m\rangle=\left\{ \begin{array}{ll}
\xi_{n,m}(\rho_0)\zeta_{l,m}(z,\phi)  & \qquad \mbox{cilindrical case} \\
\xi_{n,l}(\rho_0)\zeta_{l,m}(\theta,\phi)  & \qquad \mbox{spherical case}
\end{array}\right.
\ene

To shorten notations in what follows we drop the second subindex of
$\xi$ in the cases at hand: $\xi_{n,\cdot}\equiv\xi_n$, keeping
in mind that $n$ is a radial (rather than total) quantum number.

The operator ${\mathcal L}_d$ is considered in the space of functions
satisfying the conditions
\[
| \xi_n(0)|<\infty,\,\,\,\,\,\,\,\,\,\, \lim_{\rho_0\to
\infty}\rho_0\xi_n(\rho_0)=0\,.
\]
The inner product in this space is defined by
\bee
\label{inner_c}
(\xi_{n},\xi_{n'}) =
 \int_0^\infty\bar{\xi}_{n}(\rho)\xi_{n'}(\rho)\rho^{d-1}\,d\rho \,.
\ene
Then the operator ${\mathcal L}$ is Hermitiam.

The problem at hand is directly related to the well known eigenvalue
problem for the linear
oscillator in two and three dimensional parabolic potential \cite{LL}.
Its spectrum is discrete
\begin{equation}
\label{energy}
{\mathcal E}_{n,l,m}=
\left\{\begin{array}{ll}
4\Omega\left(n+\frac 12+\frac m2 \right)+2\Omega_z\left(l+ \frac
12\right)  & \qquad \mbox{cilindical case} \\
4\Omega\left(n+\frac 34+\frac l2 \right) & \qquad \mbox{spherical case}
\end{array}\right.
\end{equation}
(both cases naturally coincide after substitution $\Omega_z
\mapsto\Omega$ and $l+m\mapsto l$ in the first formula).
The orthormalized eigenfunctions of ${\mathcal L}_d$ are given in the
Appendix.

In accordance with the multiscale method, which is a theory for
weakly nonlinear problems,  the solution is searched in the form
of a weakly modulated linear mode. This is a main feature
 of the multiscale method, which means that the theory is applicable to {\em small condensates}.
 However, in many problems the results coming out from the
  multiscale expansions remain valid even for the large nonlinearity
  limit.
Respectively we introduce a formal small parameter $\epsilon \ll 1$ and
look for the solution of (\ref{e2}) in the form
\begin{equation}
\label{expan}
\psi= \epsilon \psi_1 + \epsilon^2 \psi_2 + ... = \left(\epsilon\xi_n^{(1)}
+\epsilon^2\xi_n^{(2)}+...\right)\zeta_{l,m}
\end{equation}

Some peculiarities of the expansion introduced are to be discussed
here. First, we use the expansion of only radial part of the solution.
This means that the consideration will be restricted to symmetric
solutions (the generalization is straightforward but cumbersome).
Second, in what follows we will concentrate on modulation of the
ground state which turns out to be flat compared with the excitation
itself. This however is not in contradiction with the assumption
about smooth (i.e. dependent on $\rho_1$) modulation of the underlying
linear mode, depending on the rapid variable $\rho_0$, by weak
nonlinearity. This is justified by the following reasons: (i) The
background solution must possess the ``right" assymptotic behavior at
$\rho\to\infty$, which can be obtained only by taking into account the
term $\Omega^2\rho^2$ at the ``linear level"; (ii) The expansion must
take into account the contribution of all linear eigenmodes, rather
than only the leading one [see (\ref{psi2}) and discussion below];
(iii) The linear problem with the parabolic trap is solved exactly.

The functions $\xi_n^{(j)}$ depend on the set of spatial coordinates
$\rho_j=\epsilon^j\rho$  and times $\tau_j=\epsilon^j \tau$
($j=0,1,2,...$): $\xi_n^{(j)}\equiv\xi_n^{(j)}
(\rho_0,\rho_1,\rho_2,...;\tau_0,\tau_1,\tau_2,...)$.
Hereafter $\rho=|\tilde{{\bf r}}_\bot|$ in the cilindrical
case and $\rho=|\tilde{{\bf r}}|$ in the spherical case.
The variables $\rho_j$ and $\tau_j$ are regarded as independent: thus
\begin{mathletters}
\begin{eqnarray}
\frac{\partial}{\partial \tau} & = &
\frac{\partial}{\partial \tau_0}+
\epsilon \frac{\partial}{\partial \tau_1}+
\epsilon^2 \frac{\partial}{\partial \tau_2}+O(\epsilon^3) \\
\frac{1}{\rho^{d-1}}\frac{\partial}{\partial \rho}\rho^{d-1}
\frac{\partial}{\partial \rho} & = &
\frac{1}{\rho_{0}^{d-1}}\frac{\partial}{\partial \rho_{0}}\rho_{0}^{d-1}
\frac{\partial}{\partial \rho_{0}}+
%\nonumber \\
\epsilon\left(2\frac{\partial^2}{\partial\rho_0\partial\rho_1}+
\frac{d-1}{\rho_0}\frac{\partial}{\partial \rho_1}\right)+O(\epsilon^2)
\end{eqnarray}
\end{mathletters}

Before going into details of the multiscale expansion we have to
clarify the physical meaning of the parameter $\epsilon$.  We are
interested in excitations against a background. Then in terms
of the dimensionless wave function
$\psi_1$ the normalization condition (\ref{norm}) leads to the estimate
$N\sim\frac{\epsilon^2V}{8\pi a_s\lambda^2}$ where $V=a_0^da_z^{3-d}$
is the  volume of the condensate which is limited by the confining
potential. These estimates imply the normalization condition
\begin{equation}
\label{norm1}
\int|\psi_1(\tilde{{\bf r}},\tau)|^2\, d\tilde{{\bf
r}}=\frac{a_0^da_z^{3-d}}{\lambda^d}
\end{equation}
(notice that in the case of cilindrical symmetry, taking $\lambda=a_s$
one obtains that the new wave function is normalized to one).
In other words the effective small parameter of the multiscale expansion
can be identified as
\begin{equation}
\label{small_par}
\epsilon=O\left(\sqrt{na_s^3}\lambda/a_s\right)=O\left(\sqrt{na_s^3}\right)
\end{equation}
 where $n=N/V$
is the density of the condensate and (\ref{lambda}) is used.
Taking into account that typically $na_s^3=O(10^{-4}\div 10^{-6})$
one concludes that $\epsilon$ ($\sim 10^{-2}\div 10^{-3}$) can
indeed be used for the multiscale expansion. We notice that the
approximation of the local kernel, $\lambda\to 0$ is evidently a
particular limit of the theory. Formally, however, it is more
convenient to keep $\epsilon$ as a small parameter without
specification and substitute it by one in the final formulas.

We are interested in the evolution of an initially unperturbed
"linear" state [say, $(n_0,l_0,m_0)$th one]. Gathering the terms of the first order
with respect to $\epsilon$ expansion we conclude that $\psi_1$ can be
represented in the form (${\mathcal E}_0={\mathcal E}_{n_0,l_0,m_0}$)
\begin{equation}
\label{psi1}
\xi_0^{(1)} = A(\rho_1,...;\tau_1,...)e^{-i {\cal
E}_0\tau_0}|n_0,l_0,m_0\rangle.
\end{equation}
Hereafter the notation $(\rho_1,...;\tau_1,...)$ is used in order to
indicate the dependence on all spatial and temporal variables, which are
slower than those written explicitely
[i.e. for example in (\ref{psi1}) the amplitude depends on $\tau_1,
\tau_2, ....$].

The solution of the equation appearing in the second order of $\epsilon$
is represented in the form of the expansion
\begin{equation}
\label{psi2}
\psi_2=\sum_{n,l,m}^{} \!\!\!^{\prime} B_{n'}(\rho_1,...;\tau_1,...)
e^{-i{\cal E}_{0}\tau_0}|n,l,m\rangle.
\end{equation}
Hereafter a prime means that the sum is computed over all arguments
such that $n\neq n_0$, $l\neq l_0$, and $m\neq m_0$.
The respective equation of the second order takes the form
\begin{equation}
\label{psi21}
\sum_{n,l,m}^{} \!\!\!^{\prime}({\cal E}_0-{\cal E}_{n,l,m})B_{n,l,m}
|n,l,m\rangle+\left[ i\frac{\partial
A_n}{\partial\tau_1}+\left(2\frac{\partial}{\partial\rho_0}
+\frac{d-1}{\rho_0}\right)\frac{\partial
A_n}{\partial\rho_1}\right]|n_0,l_0,m_0\rangle=0
\end{equation}
Applying $\langle n_0,l_0,m_0|$ to this equation (see Appendix) one arrives
at the relation $\partial A/\partial \tau_1=0$.
Here we use the property
\[
(\xi_n,\hat{p}\xi_n)=i
\int_0^\infty
\xi_n(\rho_0)\left(2\frac{\partial}{\partial\rho_0}
+\frac{d-1}{\rho_0}\right)
\xi_n (\rho_0)\rho_0^{d-1}\,d\rho_0=0
\]
with
\begin{equation}
\label{momentum}
\hat{p}=i\left(2\frac{\partial}{\partial
\rho_0}+\frac{d-1}{\rho_0}\right)
\end{equation}
being the operator of the radial component of
the (dimensionless) linear momentum. Then one concludes
that $A$ does not depend on  $\tau_1$, $A\equiv
A(\rho_1,...;\tau_2,...)$.

The coefficient $B_{n_0,l_0,m_0}$ is equal to zero identically
[otherwise it could be made zero by simple renormalization of initial
conditions for (\ref{e2})].
 In order to find other coefficients
$B_{n,l,m}$ we compute the inner product
with $\langle n_0,l_0,m_0|$. Introducing matrix elements
\begin{equation}
\label{matr} p_{n'n}=(\xi_n,\hat{p}\xi_{n'})
\end{equation}
corresponding to the transition between the states $n'$ and $n$,
originated
by the  radial component of the linear momentum considered as a
perturbation, we obtain the coefficients of the second order term
\begin{equation}
\label{psi22}
B_{n,l,m}(\rho_1,...;\tau_2,...)=i\frac{p_{n_0n}}{{\cal E}_0-{\cal
E}_{n,l,m}} \frac{\partial A}{\partial \rho_1}
\end{equation}

Finally, collecting all the terms of the third order of
$\epsilon$
and using the explicit form of the second order addendum found above we
arrive at the dynamical equation for the amplitude $A$
\begin{equation}
\label{NLS}
i\frac{\partial A}{\partial\tau_2}+ {\cal D}_0\frac{\partial^2
A}{\partial\rho_1^2}-\chi_{n_0n_0} |A|^2A=0
\end{equation}
Here
\begin{equation}
\label{disp} {\cal D}_0={\mathcal
D}_{n_0,l_0,m_0}=1-\sum_{n,l,m}^{}\!\!\!^{\prime}
\frac{|p_{n_0n}|^2}{{\cal E}_{n,l,m}-{\cal E}_0}
\end{equation}
is the effective dispersion and
\begin{equation}
\label{nonlin}
\chi_{nn'}=\int \xi_n^2(\rho)\xi_{n'}^2(\rho')
U(|\tilde{{\bf r}}-\tilde{{\bf r}}'|)
d\tilde{{\bf r}}d\tilde{{\bf r}}'
\end{equation}
is the effective nonlinearity. As a matter of fact the effective
nonlinearity can be computed under the assumption that the
characteristic scale of the ground state is much larger than the
width of the kernel $U(|\tilde{{\bf r}}|)$ of the interaction
potential. Taking into account (\ref{lambda}), (\ref{paramet}),
and (\ref{cil1}), (\ref{spher1}) one concludes that this
supposition is valid with great accuracy for several lowest
eigenstates $\xi_n(\rho)$. Then $U(|{\bf r}|)$ in the integrand of
(\ref{nonlin}) can be substituted by the delta function and thus
$\chi_{nn}=(\xi_n^2,\xi_n^2)$.

In the explicit form of ${\cal D}_0$ one can recognize the second
order addendum to the energy of an effective
particle induced by the perturbation operator
$\hat{p}$. Thus the effective dispersion has a contribution
of the  probability
of the energy transfer among the levels in the confining potential.
Notice that the mentioned transition occur among levels with the same
orbital and azimutal quantum numbers (in the case at hand they are
equal to zero).

Equation (\ref{NLS}) is considered on the semiline ($\rho_1\geq 0$)
subject to the boundary condition
$
\lim_{\rho_1\to\infty}A(\rho_1,...;\tau_2,...)={\cal A}e^{i\tilde{\omega}\tau_2}
$
where the constant ${\cal A}$ can be interpreted as an amplitude of the
fundamental
state of the condensate: after all it must be found from the
normalization condition. [Notice that other boundary conditions, of
less practical interest, are also compatible with the problem at hand].

Let us now return to the discussion of the main scales of the problem.
 In the dimensionless variables the spatial
scale of the amplitude $A(\rho_1,...;\tau_2,...)$ can be estimated as
$1/\epsilon$. On the other hand the problem possess another important
dimensionless scale $1/\sqrt{\Omega}$ which determines the wave function
localization due to the confining potential. An interesting and
physically relevant situation appears when $\epsilon\gg\sqrt{\Omega}$,
what in physical variables means $8\pi Na_s/a_0\gg 1$ \cite{com2},
a condition which is satisfied in current experimental setups.
The last requirement can be understood equivalently as smallness of
 the healing length $\xi$ compared with the linear oscillator length
$\xi\ll a_0$. In this situation the excitation width is much smaller
than the volume
``available" for the condensate and in the case of positive scattering
length one can obtain localized excitations against the background. This is
the regime of validity of our theory as has been put in more mathematical terms
in the first part of this section.

\section{Application: Vortex shape and stability}
\label{stability}

One of the simplest excitations of the ground state condensate one can construct
have the form of  static "dark" soliton solutions of Eq. (\ref{NLS}), i. e. ``black holes" or vortices (when a local phase is added). Joining the different contributions from the multiscale
method we obtain
\bee
\label{solit2d}
\Psi=\frac{1}{\sqrt{\pi}}\frac{a_s}{\xi}
\exp\left[-i\left(\frac{2}{a_0^2}+\frac{1}{a_z^2}+
8\pi \sqrt{\frac{n}{a_s}}\right)
\frac{\hbar t}{2m}\right]
\exp\left(-\frac{r_{\bot}^2}{2a_0^2}-\frac{z^2}{2a_z^2}\right)
\tanh\left(\frac{r_{\bot}}{\sqrt{2{\mathcal D}_0}\xi}\right) e^{im\phi}
 +O(na_s^3).
\ene
This solution approximates the shape of a vortex line with axis
 of rotation $z$ in trap of transverse size $a_0$.

This is not the only solution one can construct, as an example we present also a solution
corresponding to a ``black hole" in three dimensions
\bee
\label{solit3d}
\Psi=\frac{1}{\pi^{3/4}}\frac{a_s}{\xi}
\exp\left[-i\left(\frac{3}{a_0^2}
+2^{7/2}\sqrt{\frac{\pi n}{a_s}}\right)
\frac{\hbar
t}{2m}\right]\exp\left(-\frac{r^2}{2a_0^2}\right)
\tanh\left(\left(\frac{2}{\pi}\right)^{1/4}\frac{r}{\sqrt{2{\mathcal
D}_0}\xi}\right)
+O(na_s^3)
\ene
where $r=(x,y,z)$.

In the above formulas we have neglected the difference between
$\lambda$ and $a_s$ [see (\ref{lambda})]. They however still
contain the quantity ${\mathcal D}_0$ to be estimated. To this end
we employ (\ref{cil1}) and (\ref{spher1}) and compute (\ref{disp})
\cite{maple}. The result is ${\mathcal D}_0\approx 0.586$ and
${\mathcal D}_0\approx 0.728$ for (\ref{solit2d}) and
(\ref{solit3d}) respectively. Thus in the physical units the
effective width of the black hole can be estimated to be of order
of the healing lengths which is the expected result for the vortex
size.

It must be said that the result naturally does not depend on
$\Omega$, a fact which is explained by taking into account that in
the parabolic trap the energy levels are equidistant. Another
point to be mentioned is that as a matter of fact we have computed
the terms of order of $O(na_s^3)$. They are given by (\ref{psi2}),
(\ref{psi22}) and are not represented explicitly because of
cumbersome form. In other words the theory provides one with the
BEC wavefunction computed with accuracy $O((na_s^3)^{3/2})$ to
problems whose number of particles is not too large.

One of the advantages of the multiscale method is that it
guarantees the {\em dynamical stability} of the solution up to a
certain time scale. In order to explain this last point we notice
that the method is based on eliminating the secular terms in the
third order with respect to the small parameter. This in
particular means that the characteristic time of evolution of
instabilities of the solutions (if any) is given by $\epsilon^3
\tau$. The main implication is that the instabilities against the
background solution (which is the vortex in the case at hand)
cannot develop up to $\tau \sim \epsilon^{-3}$. Recalling now
(\ref{small_par}) and returning to the physical variables we
obtain that for  the typical values of $\epsilon$ $\sim$ 10$^{-2}$
any instability could develop only for times larger than 10
seconds, which is of the order of the life-time of the condensate.
In fact, this prediction is consistent with the previous numerical
observation \cite{Juanjo} that the vortex is a stable object.

 This result is relevant since a mathematical proof of the expected
dynamical stability of a vortex in a trap (i.e. when dissipation is
zero) is not available yet. However, the numerical simulations (which
are also valid only for limited times) point that a vortex should be a
 dynamically stable object. The analysis of the linearized operator
around the vortex done in Ref. \cite{Juanjo} was not completely conclusive
since there were some zero eigenvalues which could lead to instabilities
mediated by high order terms in the linearized evolution operator.
In this sense the multiscale technique adds a new argument in favor of
long time stability of
the vortex {\em in the framework of the mean field modelization of the
phenomenon}.

We remark that the previous discussion applies to {\em dynamical stability}, i. e. stability
of the solution under perturbations on the initial data. It was discussed
on the previous works on vortex stability that the system is energetically
 unstable \cite{Fetter,Juanjo}, while the question of its Lyapunov
stability remains open. Our analysis concerns only the subject of
{\em dynamical stability} and confirm the numerical observations by
many authors that vortices in a trap are very long lived and could even
 be completely stable \cite{Juanjo}.

\section{Conclusion}
\label{conclussions}

In the present paper we have developed the multiscale approach for
obtaining explicit (approximate) solutions to the GP equation for
condensates with sufficiently small number of particles which are
characterized by fast spatial scales of the order of the healing
length, and thus cannot be described within the framework of the
conventional hydrodynamical approach. One of the advantages of the
method is that it guarantees the stability of the solution as
discussed in Sec. \ref{stability} which {\em allows us to put
lower bounds on the time a vortex survives under possible
dynamical instabilities} described within the framework of the GP
equation.

Another outcome of the theory developed above is that the solutions of
the healing length size cannot be excited only as a background state.
Some accompanion modes, corresponding to higher levels of the confining
potential are excited simultaneously. All these contributions are included in
the constants ${\cal D}_0$ appearing in the reduced equations (\ref{NLS}).

One more advantage of the multiscale method is that, by reducing
the problem to solving the effective one dimensional nonlinear
Schr\"{o}dinger equation, one gets a powerful tool for obtaining
other more complicated solutions and even interactions among them,
a point which will be the subject of further study.

 The present paper provides a new analytical tool which we hope will
 be of interest for theoretical progresses in the analysis of BEC problems.

\acknowledgments

We are in debt to G. Alfimov for many fruitful discussions. V. V. K.
is grateful to University of Castilla-La Mancha for warm hospitality.
V.M.P-G. has been supported by DGCYT under grant PB96-0534.
The cooperative work has been supported through the bilateral program
 DGCyT-HP1999-019/A\c{c}\~ao  N$^o$
E-89/00.

\appendix
\section{Eigenfunctions}

For the sake of convenience here we present the orthonormalized
eigenfunctions. In the case of potential with cilindrical symmetry
they are \beq \label{cil1}
\xi_{n,m}(\rho)=\sqrt{2\Omega\frac{n!}{(n+m)!}}\left(\sqrt{\Omega}\rho\right)^m
e^{-\Omega\rho^2/2}L_n^{(m)}(\Omega\rho^2), \enq \beq \label{cil2}
\zeta_{l,m}(\phi,z)=\frac{\Omega_z^{1/4}}{\pi^{3/4}\sqrt{2^{l+1}l!}}e^{im\phi}e^{-\Omega_z
z^2/2}H_l\left(\sqrt{\Omega}_z z\right). \enq
 Here $H_n(\cdot)$ is the Hermite polinomial and $L_n^{(m)}(\cdot)$ is
the Laguerre polinomial. When spherical symmetry is imposed the eigenfunctions are given by
\bee
\label{spher1}
\xi_{n,l}(\rho)= 2
\left[\frac{2^{n+l}}{(2(n+l)+1)!!}\Gamma (n+1)\right]^{1/2}
\left(\frac{\Omega^3}{\pi}\right)^{1/4}
e^{-\Omega \rho^2/2}
(\sqrt{\Omega}\rho)^l
L_{n}^{(l+1/2)}(\Omega r^2),
\ene
\bee
\label{spher2}
\zeta_{l,m}(\theta,\phi)\equiv Y_{l}^{m}(\theta,\phi)=(-1)^{m+|m|}i^l\left[
\frac{2l+1}{4\pi}\frac{(l-|m|)!}{(l+|m|)!}\right]^{1/2}P_l^{|m|}(\cos\theta)
e^{im\phi}.
\ene

The inner product in this space is defined by
\begin{equation}
\label{inner_cc} \langle n,l,m|n',l',m'\rangle
=\int_{-\infty}^{\infty} dz\int_0^{2\pi}d\phi\int_{0}^{\infty}
d\rho \rho \bar{\xi}_{n,m}(\rho)
\xi_{n',m'}(\rho)\bar{\zeta}_{l,m}(z,\phi) \zeta_{l',m'}(z,\phi)
\end{equation}
in the case  of cilindrical symmetry and by
\begin{equation}
\label{inner_s} \langle n,l,m|n',l',m'\rangle=\int_{-\pi}^{\pi}
d\theta\int_0^{2\pi}d\phi\int_0^\infty d\rho \rho
\bar{\xi}_{n,m}(\rho)\xi_{n',m'}(\rho)\bar{\zeta}_{l,m}(\theta,\phi)
\zeta_{l',m'}(\theta,\phi)
\end{equation}
in the case  of spherical simmetry


\begin{references}
\bibitem{maple}{The computation have been made with the MAPLE's codes
for the Laguerre. In the $2d$ and $3d$ cases the number of
sum entries computed were 100. However the accuracy of oder of 1\% is
achieved by taking into account 30 first terms of both sums.}
\end{references}
\end{document}